\documentclass{article}
\usepackage{spconf,amsmath,amssymb,graphicx,booktabs,xcolor,url}
\usepackage[hidelinks]{hyperref}

\title{Region-Aware Masking for Accent-Robust Cross-Lingual Text-to-Speech}
\name{Haoqi Li, Shivam Mehta, Ravi Teja Gadde, Yinghong Lan}
\address{Netflix Inc., USA}

\begin{document}
\maketitle
\ninept

\begin{abstract}
Accent leakage remains a critical challenge in cross-lingual zero-shot              text-to-speech (TTS), where models inadvertently transfer both the speaker's        timbre and source-language accent into the target-language output. In mask-reconstruction TTS this is amplified by a train--inference mismatch: training reconstructs from same-language context, while inference conditions on cross-lingual context. We close this gap by concatenating utterances from two languages and placing the reconstruction mask relative to the language boundary, so training presents the cross-lingual prompting pattern the model encounters at inference. This requires no language IDs, accent labels, parallel same-speaker recordings, or architectural changes---only the mask geometry differs. In a listening study with native speakers, region-aware masking reduces perceived accent from 4.3 to 0.5 on a 0-5 scale against the unadapted baseline, and by 18-51\% in seen regimes and 35-55\% on unseen source languages against bilingual adaptation alone, with no measurable loss in intelligibility or naturalness. Comparing mask geometries shows that placement, not the amount masked, governs the balance between accent suppression and speaker preservation: masks confined to the reconstruction region suppress accent most but retain the reference speaker least reliably, whereas two-region masking achieves both. Mask geometry is thus a simple, effective lever for accent robustness.

\end{abstract}

\vspace{-1mm}
\section{Introduction}
\label{sec:intro}
\vspace{-1mm}
Cross-lingual zero-shot text-to-speech (TTS) often exhibits accent leakage:
although the synthesized speech preserves the reference speaker's timbre, its
target-language pronunciation commonly inherits the source-language accent.
This issue arises in voice-preserving speech translation, media localization,
and multilingual voice interfaces, where speaker characteristics should remain
stable while pronunciation follows target-language norms. However, addressing
this challenge through direct supervised learning is difficult, as parallel
cross-lingual speech from the same speaker is scarce and expensive to collect
at scale.

The problem stems from two related sources. First, speaker timbre is
entangled with language-dependent pronunciation and prosody, allowing a model
to encode them jointly. Prior work addresses this using language embeddings
or tokens, disentangled speaker and linguistic representations, adversarial
accent suppression, language-dependent generators, or inference-time model
fusion \cite{vallex,dsetts,crossspeechpp,accentbox,crossaccent,xvoice,pflux}.
While these methods highlight the need to separate speaker-related cues from
pronunciation, they leave open a complementary question: can the training
objective itself provide an effective separation signal without explicit
language IDs, accent, or other token- or embedding-based conditioning?

Second, mask-reconstructing TTS introduces a fundamental train--inference
mismatch. Voicebox \cite{voicebox}, E2 TTS \cite{e2tts}, and F5-TTS
\cite{f5tts} learn speech infilling by masking part of an utterance and
reconstructing it from text and surrounding audio. During standard training,
both the masked target and its acoustic context come from the same language
and utterance. During cross-lingual inference, however, target speech in one
language must be generated using acoustic context from another. This mismatch
encourages the model to copy language-dependent acoustic attributes---most
notably, the source-language accent---from the reference audio rather than
relying on the target text for pronunciation.

In this work, we propose a region-aware masking method for accent-robust
cross-lingual TTS. Building on F5-TTS \cite{f5tts}, we construct
pseudo-bilingual examples by concatenating utterances from two distinct
languages, optionally guided by speaker-embedding similarity. By
reconstructing speech in one language while conditioning on the other as
acoustic context, the model is exposed during training to the same
cross-lingual prompting pattern encountered at inference time. This directly
targets the train--inference mismatch underlying accent leakage, without
explicit language IDs, accent conditioning, or architectural modifications.

To summarize, our main contributions are:
\begin{itemize}
    \vspace{-0.5mm}
    \item \textbf{A region-aware cross-lingual adaptation strategy:} We
    construct pseudo-bilingual training pairs from speaker-disjoint
    monolingual data via region-aware masking, directly addressing the
    train--inference mismatch.

    \vspace{-0.5mm}
    \item \textbf{Conditioning-free accent control:} Our approach improves
    accent robustness without requiring explicit language IDs, accent
    embeddings, or model modifications.

    \vspace{-0.5mm}
    \item \textbf{A practical trade-off analysis:} We show that mask
    placement governs the trade-off between accent suppression and
    speaker-timbre preservation, identify two-region masking as the best
    overall setting, and demonstrate that the resulting gains generalize to
    unseen language combinations.
\end{itemize}

Audio demos are provided on the supplementary
website.\footnote{\url{https://doubleblindauthor.github.io/region-aware-masking-TTS-demo}}
\vspace{-1mm}
\section{Related work}
\vspace{-1mm}
\label{sec:related}

\textbf{Masked and cross-lingual generation.}
Prior work adapts masked reconstruction or codec-based language
modeling for cross-lingual speech generation. Voicebox \cite{voicebox} introduced large-scale
text-guided flow-matching infilling, and E2 TTS \cite{e2tts} and F5-TTS \cite{f5tts} simplified it
to alignment-free zero-shot TTS
. Cross-Lingual F5-TTS extends base F5-TTS toward
language-agnostic voice cloning, dropping the reference
transcript via forced alignment and speaking-rate prediction
\cite{crosslingualf5}. VALL-E established neural-codec language modeling for
zero-shot TTS \cite{valle}, and VALL-E X concatenates source- and
target-language phoneme prompts within a codec language model for
cross-lingual synthesis \cite{vallex}. SpeechPainter \cite{speechpainter} and ERNIE-SAT \cite{erniesat} apply masked reconstruction directly to speech editing and synthesis. Separately, teacher-student augmentation
synthesizes unseen speaker-language combinations for cross-lingual training
\cite{teacherstudent}.

\textbf{Speaker, language, and accent separation.}
A separate line of work addresses accent leakage by separating speaker
cues from linguistic content. Zhang \emph{et al.} use adversarial training to
separate the two and show that accented outputs can still receive higher
speaker-similarity ratings \cite{foreignlanguage}, underscoring that
similarity metrics alone cannot certify accent-free synthesis. YourTTS \cite{yourtts} and XTTS \cite{xtts} extend zero-shot synthesis across languages .
DSE-TTS separates a native-speaker representation for linguistic style from
target-speaker timbre \cite{dsetts}; CrossSpeech++ uses separate language-
and speaker-dependent generators \cite{crossspeechpp}; AccentBox and
CrossAccent-TTS learn explicit accent representations or suppression modules
\cite{accentbox,crossaccent}; X-Voice injects language identity at the text
and frame levels \cite{xvoice}; PFluxTTS applies inference-time
vector-field fusion \cite{pflux}. Accent normalization has also been studied
using non-parallel data and a separate discrete-token conversion model
\cite{accentnorm}.

Both lines rely on masked or codec-based reconstruction with explicit
language, accent, or speaker representations. We instead study how mask
placement relative to the language boundary changes the training signal itself,
improving accent robustness without language IDs, accent labels, or
architectural changes.
\vspace{-1mm}
\section{Method}
\label{sec:method}
\vspace{-2mm}
\subsection{Masked flow-matching supervision}
\label{ssec:cfm}

Our method builds on F5-TTS \cite{f5tts}, a non-autoregressive
diffusion-transformer TTS model trained for text-guided speech infilling. It
uses optimal-transport conditional flow matching (OT-CFM) to learn a vector
field between Gaussian noise and speech data \cite{lipman,cfm}. Let
$x_1\in\mathbb{R}^{T\times F}$ be a log-mel spectrogram and $y$ its transcript.
OT-CFM samples $x_0\sim\mathcal{N}(0,I)$ and $t\sim\mathcal{U}[0,1]$, and
defines $\psi_t=(1-t)x_0+tx_1$ and $\dot{\psi}_t=x_1-x_0$. Let
$m\in\{0,1\}^{T}$ indicate frames to reconstruct and
$c=(1-m)\odot x_1$ be the visible acoustic context. The masked CFM objective is
\vspace{-2mm}
\begin{equation}
\label{eq:loss}
\mathcal{L}_{\mathrm{MCFM}}=
\mathbb{E}_{\substack{t,\,x_1\sim q,\,x_0\sim p_0,\\m\sim p_{\mathrm{mask}}}}
\left\|m\odot\left(v_\theta(\psi_t,t,c,y)-\dot{\psi}_t\right)\right\|_2^2
\end{equation}
Here, $v_\theta$ is the time-dependent vector field predicted by F5-TTS, and
$q$, $p_0$, and $p_{\mathrm{mask}}$ denote the speech-data, Gaussian-noise, and
training-mask distributions. Eq.~\eqref{eq:loss} omits the
normalization factor, as in prior work; in practice, we average over masked
time--frequency bins. Standard F5-TTS masks one random contiguous span of
length $sT$, where $s\sim\mathcal{U}[0.7,1.0]$ \cite{f5tts}.
\vspace{-2mm}
\subsection{Construction of concatenated bilingual examples}
\label{ssec:concat}

Let $\mathcal{D}_0$ and $\mathcal{D}_1$ denote monolingual corpora for
Language~0 and Language~1. We sample one utterance from each corpus,
concatenate their log-mel spectrograms, and join their character-token
sequences with the in-vocabulary separator ``\texttt{|}''. The concatenation
order is sampled uniformly, exposing the model to both language directions.
After ordering, we refer to the first and second positional regions as
\emph{part0} and \emph{part1}; these names do not correspond to a fixed
language. Let their durations be $T_0$ and $T_1$ mel frames.

The language boundary determines how the example is masked, but no language ID
is passed to the model. The two utterances need not be semantically aligned or
recorded from the same speaker. To preserve ordinary monolingual behavior, a
fraction $p_{\mathrm{single}}$ of training examples remains monolingual and
uses the standard F5-TTS mask; the remaining examples use the concatenated
construction. At inference, the reference audio is prompted with its
transcript and the target text joined by the same separator, and the model
generates the target-language continuation.

\begin{figure}[!t]
\centering
\includegraphics[width=0.8\columnwidth]{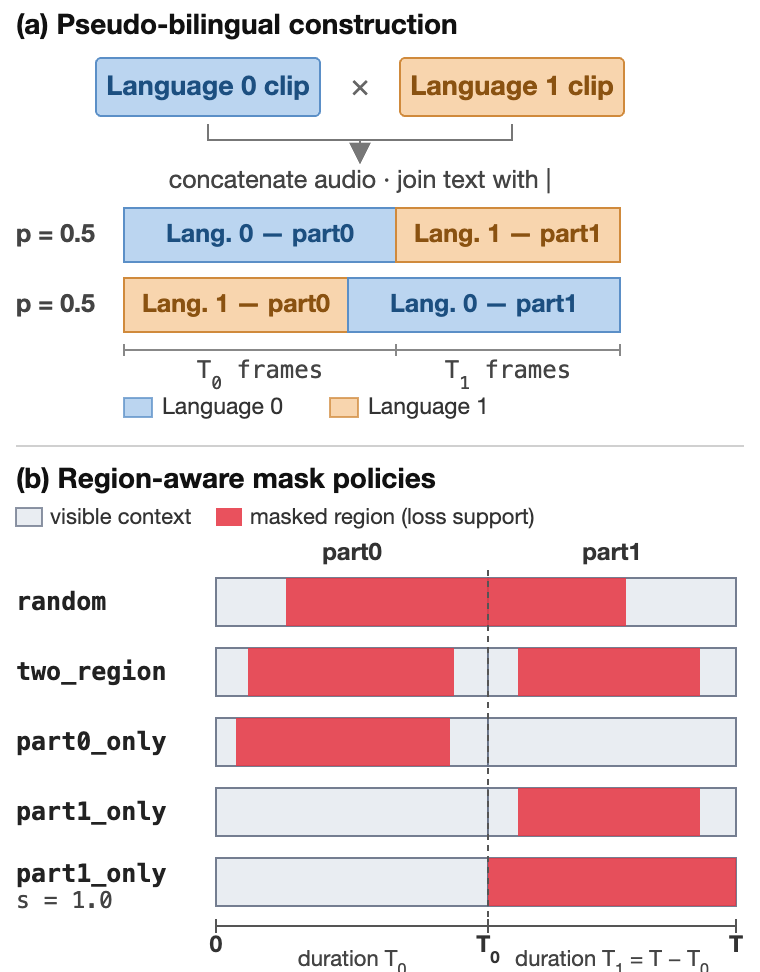}
\caption{\footnotesize Pseudo-bilingual example construction and region-aware masking.
(a) Two utterances are concatenated into regions of durations $T_0$ and $T_1$.
(b) Gray denotes visible acoustic context and red denotes masked regions used
in the flow-matching loss.}
\label{fig:masks}
\vspace{-5mm}
\end{figure}

\vspace{-3mm}
\subsection{Region-aware mask distributions}
\label{ssec:masks}

Our key idea is to place two languages in one training example, then mask one
region so the model reconstructs it while receiving the other as acoustic context. This encourages transfer of language-independent cues such as timbre across the boundary while relying on the target text for pronunciation. The original random-span mask does not reliably create this condition, since it may cross the language boundary or leave target-region speech visible. We therefore define several position-aware masking strategies.


For an interval beginning at $t_{\mathrm{start}}$ and ending at
$t_{\mathrm{end}}$, let
$\mathrm{span}(t_{\mathrm{start}},t_{\mathrm{end}},s)$ denote a uniformly
positioned contiguous mask of duration
$\lfloor s(t_{\mathrm{end}}-t_{\mathrm{start}})\rfloor$, with masked
fraction $s$. With $T=T_0+T_1$, we define four policies: \emph{random}
masks a span in $[0,T]$; \emph{part0-only} masks a span in $[0,T_0]$;
\emph{part1-only} masks a span in $[T_0,T]$; and \emph{two-region}
independently masks one span in each region. Formally, these correspond to
$\mathrm{span}(0,T,s)$, $\mathrm{span}(0,T_0,s)$,
$\mathrm{span}(T_0,T,s)$, and
$\mathrm{span}(0,T_0,s_0)\cup \mathrm{span}(T_0,T,s_1)$, where
$s,s_0,s_1\stackrel{\mathrm{iid}}{\sim}\mathcal{U}[0.7,1.0]$ unless specified
otherwise. The \emph{uniform} policy samples one of the four with equal
probability at each optimization step. Because concatenation order is
randomized, \emph{part0-only} and \emph{part1-only} are symmetric over
training.

The most inference-aligned policy is \emph{part1-only}. Under the default
span-fraction distribution, part of the second region remains visible; setting
$s=1$ masks it completely, requiring the model to reconstruct it from its
transcript and cross-language acoustic context. With order randomized, either
language can occupy the reconstruction region. This signal is intended to
encourage, rather than guarantee, separation of timbre from
language-dependent pronunciation.

\subsection{Speaker-similarity-selected pseudo-pairs}
\label{ssec:pairing}
Region-aware masking does not require any cross-lingual speaker matching: the simplest
setting concatenates utterances from the two languages at random. Random
concatenation, however, can pair very different voices, weakening cross-region
acoustic context for speaker timbre. We therefore also study a pseudo-pairing
strategy based on speaker similarity.

For each utterance in an anchor language, we retrieve the top $K$ candidates
from a candidate language by cosine similarity between normalized pretrained
speaker embeddings and sample one candidate during training. This raises similarity in
speaker characteristics only; it does not guarantee that the paired utterances
share the same speaker. We therefore refer to them as \emph{pseudo-pairs}.

To quantify how strongly an embedding model separates retrieved from random
pairs, we define
\vspace{-2mm}
\begin{equation}
\rho=\frac{\mathbb{E}_i[\cos(e_i,e_{\mathrm{NN}_1(i)})]}
          {\mathbb{E}_{i,j}[\cos(e_i,e_j)]}
\label{eq:ratio}
\end{equation}
where $e_i$ is an anchor-language embedding, $e_j$ a randomly selected
candidate-language embedding, and $e_{\mathrm{NN}_1(i)}$ the nearest
candidate-language embedding to $e_i$. A value near one indicates little
separation between retrieved and random pairs under that encoder. This ratio
is descriptive rather than a direct measure of pair quality: speaker
embeddings may also encode channel, language, accent, or other attributes, and
$\rho$ evaluates the nearest candidate although training samples from the top
$K$.

\vspace{-2mm}
\section{Experimental setup}
\label{sec:setup}
\vspace{-0.5mm}

\textbf{Datasets.}
We instantiate Language~0 and Language~1 as English and Spanish. Both
languages offer large, diverse corpora, and their phonological and prosodic
differences provide a meaningful test of cross-lingual accent transfer. We
construct pseudo-bilingual training examples from an internal corpus of
approximately 21.2M English and 9.3M Spanish utterances. Each utterance is
a single-speaker audio clip with durations between 2 and 16 seconds. The two language
subsets are speaker-disjoint, and no same-speaker bilingual recordings are
used.

For evaluation, we use an internal ES$\rightarrow$EN test set of 20 verified
cross-lingual speech--text pairs, with Spanish reference speech and English
target text. To assess accent robustness beyond the training language pair and
domain, we also build a test set from FLEURS \cite{conneau2023fleurs}, which
provides semantically parallel speech and text across languages. We select 30
sentences and pair their Spanish, French, and Portuguese recordings with the
corresponding English target text, yielding 90 cross-lingual items across
three source languages. All paired utterances preserve the same semantic
content.

\textbf{Cross-lingual speaker-pair retrieval.}
For the optional pairing strategy, we precompute cross-language
nearest-neighbor tables with FAISS
\cite{douze2024faiss}, avoiding online similarity search during training.
English utterances serve as anchors, Spanish utterances as candidates. For
each anchor, we retrieve the top $K=10$ candidates by cosine similarity between
normalized speaker embeddings and sample one uniformly from this set, balancing
speaker similarity with pair diversity. After pairing, the concatenation order
during training is sampled uniformly. Table~\ref{tab:encoders} reports the
top-1 retrieval contrast from Eq.~\eqref{eq:ratio} over 1{,}000 anchors.
ECAPA-TDNN yields a substantially larger retrieved-versus-random contrast than
x-vector, indicating stronger discrimination between retrieved and random pairs
and motivating its use in the optional speaker-pairing experiments.

\begin{table}[!t]
\vspace{-1mm}
\centering
\caption{Top-1 speaker-pair retrieval statistics.}
\label{tab:encoders}
\footnotesize
\renewcommand{\arraystretch}{0.95}
\setlength{\tabcolsep}{3pt}
\begin{tabular}{@{}lcccc@{}}
\toprule
Encoder & Dim. & NN cosine & Random cosine & $\rho$ \\
\midrule
x-vector \cite{xvector} & 512 & 0.982 & 0.923 & 1.06 \\
ECAPA-TDNN \cite{ecapa} & 192 & 0.762 & 0.322 & \textbf{2.37} \\
\bottomrule
\end{tabular}
\end{table}

\textbf{Model setup.}
We adopt two measures to avoid degrading monolingual generation. First, all
systems are initialized from the same pretrained multilingual F5-TTS checkpoint, rather
than training from scratch: we evaluate it directly as baseline B and adapt the
rest from it. Second, alongside concatenated bilingual examples, we retain a
small proportion of monolingual examples during adaptation to mitigate catastrophic forgetting.

We set $p_{\mathrm{single}}=0.2$, so 80\% of
adaptation examples are concatenated and 20\% remain monolingual. During
training, we use AdamW with a learning rate of $10^{-4}$ and a 4{,}000-update
warmup. All systems are compared at the same training budget of 2.04 epochs on
the concatenated mixture, so differences reflect the method rather than
training length.

We evaluate the masking methods summarized in Table~\ref{tab:systems}. In the
Mask column, \emph{uniform} denotes a mixture of all four masking policies:
At each training step, one of the policies defined in Section~\ref{ssec:masks} is
sampled with equal probability. $U$ indicates that the corresponding mask
fraction is sampled from $\mathcal{U}[0.7,1.0]$. M+P denotes the corresponding
masking policy with the optional ECAPA-TDNN-based speaker-pair retrieval.

\begin{table}[!t]
\vspace{-6.6mm}
\centering
\caption{Systems. $U=\mathcal{U}[.7,1]$.}
\label{tab:systems}
\footnotesize
\renewcommand{\arraystretch}{0.95}
\begin{tabular}{@{}lll@{}}
\toprule
System & Mask ($s$) & Pairing \\
\midrule
B    & random ($U$)     & None (single-language utterance) \\
M1   & random ($U$)     & Random \\
M2   & two\_region ($U$)& Random \\
M3   & part1\_only ($U$)& Random \\
M4   & part1\_only ($1$)& Random \\
M5   & uniform ($U$)    & Random \\
M5+P & uniform ($U$)    & ECAPA-TDNN \\
M2+P & two\_region ($U$)& ECAPA-TDNN \\
\bottomrule
\end{tabular}
\vspace{-5mm}
\end{table}

\textbf{Evaluation.}
We report three objective measures, along with a subjective listening test. Speaker similarity (SIM) is computed using both WavLM
\cite{wavlm} and ECAPA-TDNN \cite{ecapa}. Since ECAPA-TDNN is also used for
pseudo-pair retrieval, WavLM provides an independent check against potential
bias from pair selection. Intelligibility is measured by word error rate (WER)
using Whisper large-v3 \cite{whisper}, and naturalness is estimated with UTMOS
\cite{utmos}. We treat UTMOS as a proxy for naturalness and overall generation
quality.

To directly assess accent robustness, we conduct a human listening test. Each
page presents a non-English reference utterance, the English target text, and
the generated English speech. Twenty English native speakers rate each clip on
two criteria. The first rates the perceived strength of the source-language
accent on a 0--5 scale: 0 none, 1 very slight, 2 mild, 3 moderate, 4 strong,
and 5 very strong; lower is better. The second asks whether the speaker timbre
is noticeably changed relative to the reference, which we report as the
fraction of clips flagged.

\vspace{-3mm}
\section{Results and discussion}
\label{sec:results}
\vspace{-1mm}

We report results in three configurations that cross two axes of shift: source
language and audio domain. \textbf{Internal ES$\rightarrow$EN} is in-domain on
both. \textbf{FLEURS ES$\rightarrow$EN} keeps the language pair and switches
to out-of-domain (OOD) audio. \textbf{FLEURS FR/PT$\rightarrow$EN} shifts both
at once, with French and Portuguese unseen during adaptation and retrieval.
The first two test the adapted model's language-pair performance across domains; the third evaluates generalization to unseen source languages.
In the listening
study, 20 raters score each clip for accent on a 0--5 scale and judge whether
its timbre is preserved. Inter-rater agreement is high for accent
(Krippendorff's $\alpha=0.98$) and moderate for timbre ($\alpha=0.53$), so we
interpret the latter mainly at the level of large between-policy differences.


\textbf{Region-aware masking consistently reduces accent leakage.}
The unadapted baseline \textbf{B} exhibits strong accent leakage across all three
regimes, with accent scores of 4.21--4.37, showing that cross-lingual
prompting largely transfers the source-language accent together with the
reference speaker's timbre. Adapting the pretrained multilingual checkpoint on
bilingual concatenations already mitigates this effect substantially:
\textbf{M1} reduces the accent score to 0.45 on the internal set, 0.69 on
FLEURS ES$\rightarrow$EN, and 1.05 on FLEURS FR/PT$\rightarrow$EN. All
region-aware masking variants further improve over \textbf{M1}.

Compared to \textbf{M1}, region-aware masking reduces perceived accent by
18--51\% in the seen regimes and 35--55\% on unseen source languages. These
gains are significant in all settings except \textbf{M5} on the 20-item
internal set. The largest gains appear in the hardest regime,
\textbf{FLEURS FR/PT$\rightarrow$EN}, where adaptation provides no direct
support for the source languages: the accent score drops from
1.05 for \textbf{M1} to 0.69 for \textbf{M2}, 0.60 for \textbf{M5}, and
0.48--0.54 for \textbf{M3/M4}. The trend holds in-domain and
out-of-domain, suggesting that the gain comes from the training strategy, not the recording domain.

\textbf{Accent suppression and speaker preservation trade off.}
The gains from region-aware masking come with a clear trade-off, but it is
driven by where the mask falls, not by how much it covers: \textbf{random}
and \textbf{two-region} both mask 85\% of the example in expectation, yet
spreading the mask across both regions lowers accent and raises \textbf{Joint}.
The \textbf{part1-only} variants mask only 42.5\% in expectation, yet preserve
timbre least reliably.
On unseen languages, accent
improves from 0.69 with \textbf{M2} to 0.48 with \textbf{M3}, but the share
of clips that deviate from the reference speaker's timbre rises
sharply from 15.1\% to 42.1\%. SIM exhibits the same trend: the part1-only variants show the largest
similarity drop. In several such cases the change is
perceptually obvious, indicating that these failures are substantial rather
than marginal.

This trade-off makes accent score alone an inadequate selection criterion. If
systems are ranked only by accent suppression, these part1-only variants
appear best; but they also fail most often to preserve the reference
speaker's timbre, making them unsuitable for voice-preserving applications.
To capture both requirements, we further report \textbf{Joint}: the fraction
of clips with both accent score $\leq 1$ and preserved speaker timbre. This
threshold corresponds to no accent or only a very slight accent. Under this
criterion the ranking reverses: on unseen languages, \textbf{M2}, \textbf{M2+P}, \textbf{M5} and
\textbf{M5+P} are tied-best on Joint (66--70\%), while \textbf{M3} and
\textbf{M4} drop to 51.3\% and 42.5\%. We therefore recommend
\textbf{two-region} masking as the default.

\textbf{Intelligibility and naturalness are largely preserved.}
Accent-robustness gains bring no measurable loss in intelligibility or
naturalness. Across the adapted systems, WER and UTMOS differences are generally not statistically resolvable in the seen configurations; UTMOS separates the unadapted baseline in both, while WER does so only on FLEURS.
This suggests the proposed adaptation mainly changes accent and speaker-preservation
behavior while leaving overall generation quality largely intact. The only
notable exception is \textbf{M5} on FLEURS FR/PT$\rightarrow$EN, which
improves over \textbf{M1} in WER; no other masking variant shows this
pattern.

\begin{table}[!t]
\vspace{-1mm}
\centering
\caption{\footnotesize Objective and subjective results by evaluation regime. Joint is the
share of clips with accent $\leq 1$ and preserved speaker timbre. SIM reports
WavLM~/~ECAPA-TDNN. Gray rows (B, M1) are reference points; bold marks the
tied-best proposed systems (paired bootstrap, $p>0.05$). WER is unbolded
in-domain, where no adapted systems are statistically distinguishable.}
\label{tab:alt}
\footnotesize
\setlength{\tabcolsep}{2.2pt}
\resizebox{\columnwidth}{!}{%
\begin{tabular}{@{}lcccccc@{}}
\toprule
System & Accent$\downarrow$ & Timbre chg.\ \%$\downarrow$ & Joint \%$\uparrow$
& SIM$\uparrow$ (WavLM / ECAPA) & WER$\downarrow$ & UTMOS$\uparrow$ \\
\midrule
\multicolumn{7}{c}{\emph{In-domain} --- Internal ES$\rightarrow$EN} \\
\midrule
\textcolor{gray!60!black}{B} & \textcolor{gray!60!black}{4.28} & \textcolor{gray!60!black}{1.5} & \textcolor{gray!60!black}{1.3} & \textcolor{gray!60!black}{0.953 / 0.701} & \textcolor{gray!60!black}{0.051} & \textcolor{gray!60!black}{2.88} \\
\textcolor{gray!60!black}{M1} & \textcolor{gray!60!black}{0.45} & \textcolor{gray!60!black}{10.8} & \textcolor{gray!60!black}{79.1} & \textcolor{gray!60!black}{0.673 / 0.313} & \textcolor{gray!60!black}{0.042} & \textcolor{gray!60!black}{3.77} \\
M2 & 0.32 & \textbf{14.7} & \textbf{80.2} & \textbf{0.645 / 0.249} & 0.060 & \textbf{3.85} \\
M3 & \textbf{0.22} & 45.3 & 52.3 & 0.520 / 0.079 & 0.024 & \textbf{4.01} \\
M4 & \textbf{0.26} & 51.0 & 45.2 & 0.478 / 0.066 & 0.039 & \textbf{3.93} \\
M5 & 0.37 & 21.3 & 71.1 & 0.614 / 0.198 & 0.036 & 3.86 \\
M5+P & \textbf{0.23} & 26.5 & 70.6 & 0.578 / 0.171 & 0.054 & \textbf{3.91} \\
M2+P & 0.36 & \textbf{11.4} & \textbf{82.4} & \textbf{0.653 / 0.261} & 0.042 & \textbf{3.95} \\
\midrule
\multicolumn{7}{c}{\emph{OOD, seen source language} --- FLEURS ES$\rightarrow$EN} \\
\midrule
\textcolor{gray!60!black}{B} & \textcolor{gray!60!black}{4.37} & \textcolor{gray!60!black}{1.4} & \textcolor{gray!60!black}{0.7} & \textcolor{gray!60!black}{0.951 / 0.717} & \textcolor{gray!60!black}{0.070} & \textcolor{gray!60!black}{3.38} \\
\textcolor{gray!60!black}{M1} & \textcolor{gray!60!black}{0.69} & \textcolor{gray!60!black}{7.4} & \textcolor{gray!60!black}{75.3} & \textcolor{gray!60!black}{0.722 / 0.406} & \textcolor{gray!60!black}{0.049} & \textcolor{gray!60!black}{3.97} \\
M2 & 0.50 & \textbf{12.7} & \textbf{75.9} & 0.671 / 0.270 & 0.049 & \textbf{3.90} \\
M3 & \textbf{0.45} & 55.0 & 40.1 & 0.507 / 0.073 & 0.037 & \textbf{3.88} \\
M4 & \textbf{0.36} & 73.5 & 23.4 & 0.427 / 0.069 & 0.037 & \textbf{4.00} \\
M5 & \textbf{0.45} & 29.2 & 62.4 & 0.623 / 0.218 & 0.058 & 3.86 \\
M5+P & \textbf{0.40} & 25.2 & 66.9 & 0.603 / 0.224 & 0.058 & \textbf{3.95} \\
M2+P & 0.52 & \textbf{10.0} & \textbf{77.4} & \textbf{0.699 / 0.316} & 0.055 & \textbf{4.04} \\
\midrule
\multicolumn{7}{c}{\emph{OOD, unseen source languages} --- FLEURS FR/PT$\rightarrow$EN} \\
\midrule
\textcolor{gray!60!black}{B} & \textcolor{gray!60!black}{4.21} & \textcolor{gray!60!black}{1.5} & \textcolor{gray!60!black}{3.1} & \textcolor{gray!60!black}{0.960 / 0.710} & \textcolor{gray!60!black}{0.122} & \textcolor{gray!60!black}{3.20} \\
\textcolor{gray!60!black}{M1} & \textcolor{gray!60!black}{1.05} & \textcolor{gray!60!black}{10.0} & \textcolor{gray!60!black}{64.5} & \textcolor{gray!60!black}{0.776 / 0.358} & \textcolor{gray!60!black}{0.089} & \textcolor{gray!60!black}{3.81} \\
M2 & 0.69 & \textbf{15.1} & \textbf{69.6} & 0.704 / 0.234 & 0.066 & \textbf{3.94} \\
M3 & \textbf{0.48} & 42.1 & 51.3 & 0.558 / 0.072 & 0.064 & \textbf{3.89} \\
M4 & \textbf{0.54} & 52.2 & 42.5 & 0.535 / 0.053 & 0.066 & \textbf{3.85} \\
M5 & 0.60 & 22.0 & \textbf{65.7} & 0.663 / 0.168 & \textbf{0.035} & \textbf{3.93} \\
M5+P & 0.62 & \textbf{16.2} & \textbf{69.4} & 0.694 / 0.241 & \textbf{0.055} & \textbf{3.87} \\
M2+P & 0.67 & \textbf{16.0} & \textbf{69.7} & \textbf{0.718 / 0.277} & \textbf{0.038} & \textbf{3.92} \\
\bottomrule
\end{tabular}%
}
\vspace{-6mm}
\end{table}

\textbf{Speaker-similarity pairing is an optional extension that mainly improves SIM under aggressive masking.}
We include pseudo-pair retrieval because purely random cross-lingual
concatenation can pair very different voices. Its effect is smaller and more
selective than that of masking itself. Under \textbf{two-region} masking, some
same-speaker acoustic context remains visible, so retrieval yields only modest
SIM gains (+0.046 on FLEURS ES and +0.043 on unseen languages). Under
\textbf{uniform} masking, the target region can be fully masked, leaving no
same-speaker evidence; the retrieved pair then becomes the only speaker anchor
and yields the largest gain we observe (+0.073 on unseen languages), 
while it is small on FLEURS ES (+0.006)
and slightly negative in-domain.
Because ECAPA-TDNN is used both for
retrieval and as one similarity metric, we also report WavLM-based SIM as an
independent check. The gains persist there at smaller magnitude. However,
these objective improvements do not translate into a clearly resolved gain in
human timbre judgments, and pairing does not materially affect accent.

We therefore view pairing as optional: it can improve similarity when masking
removes most speaker evidence, but the main gains in accent robustness come
from region-aware masking. More broadly, SIM should be interpreted carefully in
cross-lingual accent evaluation, since higher SIM can also reflect stronger
source-accent copying. This is most evident in baseline \textbf{B}, which
achieves the highest SIM while also exhibiting the strongest accent leakage.
Taken together, these results identify \textbf{two-region} masking as the best
trade-off between accent robustness and speaker-timbre preservation.

\vspace{-2mm}
\section{Conclusion}
\label{sec:conclusion}
\vspace{-2mm}
We introduced a region-aware masking strategy for accent-robust cross-lingual
TTS. By reconstructing speech in one language from acoustic context in
another, the method better aligns masked training with cross-lingual
inference, without architectural changes, explicit language or accent
conditioning, or same-speaker bilingual recordings.
Experiments show that this change to the training signal consistently reduces
perceived accent, generalizes to unseen source languages, and reveals a clear
trade-off between accent suppression and speaker-timbre preservation. Mask
geometry thus provides an effective lever for controlling accent robustness.


\end{document}